\documentclass[
,aps
 ,twocolumn
 ,secnumarabic
,amssymb, amsmath,nobibnotes, aps, prl, floatfix]{revtex4-1}

\expandafter\ifx\csname package@font\endcsname\relax\else
 \expandafter\expandafter
 \expandafter\usepackage
 \expandafter\expandafter
 \expandafter{\csname package@font\endcsname}%
\fi

\def\BR{\mbox{\rm BR}}

\usepackage{epsfig}
\usepackage{hyperref}
\usepackage{docs}
\usepackage{bm}

\begin{document}

\title{Searches for Majorana Neutrinos and Direct Searches for Exotics at LHCb}
\author{Xabier Cid Vidal}
\affiliation{European Organization for Nuclear Research\\
CH-1211 Geneva 23\\
Switzerland}
\collaboration{on behalf of the LHCb collaboration}
\date{17 October 2015}

\begin{abstract}
These proceedings present the LHCb results on Majorana neutrino searches and direct production of exotic particles using the data collected during Run I of LHC. For the former, Majorana neutrinos are searched for both on-shell and off-shell in $B$ and $D$ decays to final states with two same-sign muons. For the latter, different types of new particles are studied profiting the unique coverage of LHCb with respect to other detectors.
\end{abstract}

\thispagestyle{empty}
\maketitle

\section{Introduction}

\begin{figure}[ht]

\vspace{1cm}
\includegraphics[width=0.45\textwidth]{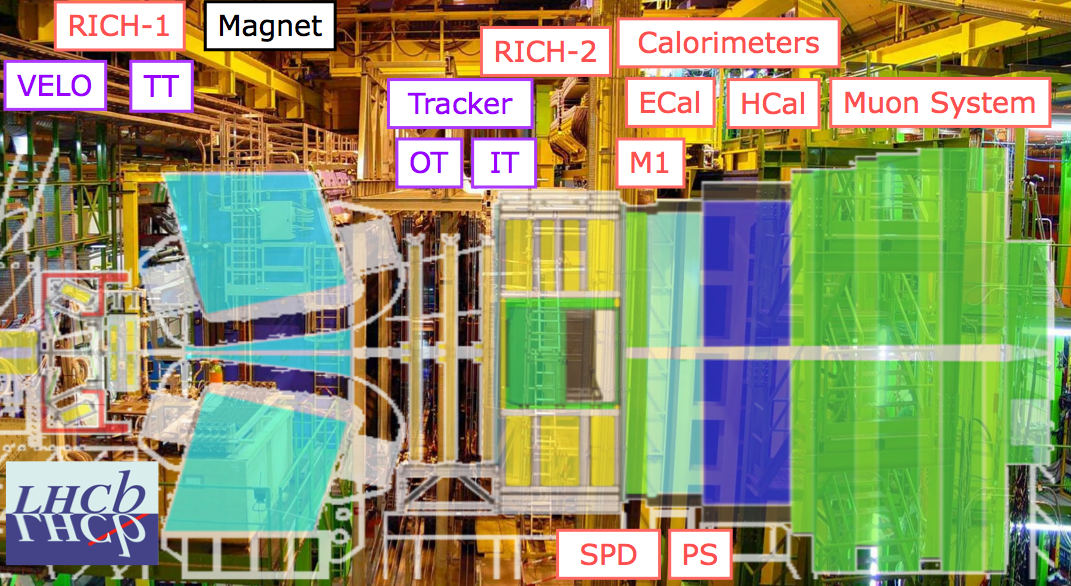}

\vspace{.5cm}
\hspace{0.5cm}\includegraphics[width=0.08\textwidth]{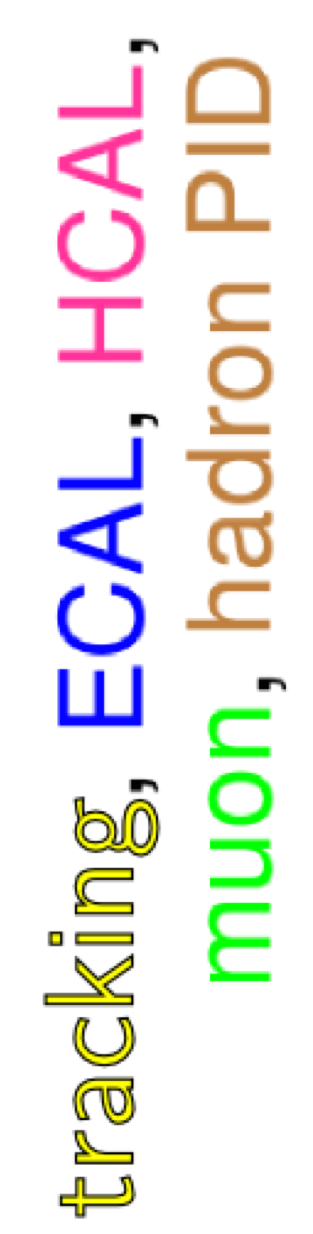}
\includegraphics[width=0.25\textwidth]{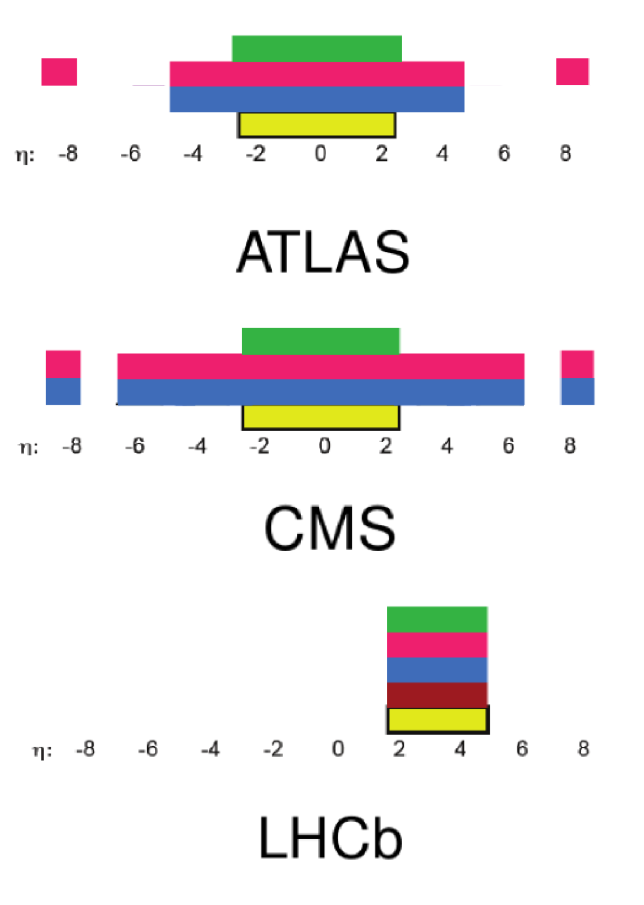}
\caption{Top: Photograph of the LHCb detector with its subdetectors highlighted. Bottom: $\eta$ coverage comparison between LHCb, ATLAS and CMS.}
\label{fig:lhcb}
\end{figure}

LHCb is one of the four primary detectors of the Large Hadron Collider (LHC) project at CERN (Geneva, Switzerland). LHCb is a single-arm forward spectrometer with a unique coverage in terms of pseudorapidity, $\eta$ (see FIG. \ref{fig:lhcb}), originally designed to study the production and decay of $b-$ and $c-$hadrons but currently extending its physics programme to include also other areas such as electroweak or exotica searches. Exotica at LHCb includes both the direct production of particles beyond the Standard Model and Higgs physics. The strong points of  LHCb are an excellent particle identification (including $K /\pi$ separation), secondary vertex, lifetime, momentum and invariant mass resolution.

During Run I of LHC, LHCb recorded data at centre of mass energies of $\sqrt{s}=7$ TeV (2010 and 2011) and 8 TeV (2012). To optimise the performance for $b-$physics, LHCb runs with lower instantaneous luminosity than ATLAS and CMS, with the advantage of having stable conditions in terms of the instantaneous luminosity during a fill (thanks to the luminosity levelling). LHCb took 37 pb$^{-1}$ in 2010, 1 fb$^{-1}$ in 2011 and 2 fb$^{-1}$ in 2012, respectively. 

\begin{figure}[htbp]
\includegraphics[width=0.45\textwidth]{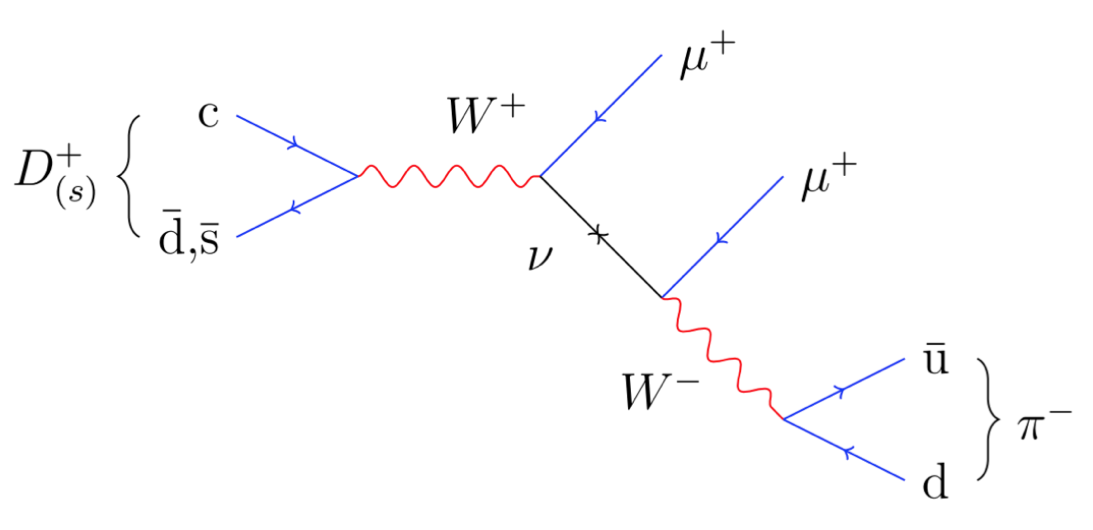}

\vspace{.5cm}
\hspace{0.5cm} \includegraphics[width=0.45\textwidth]{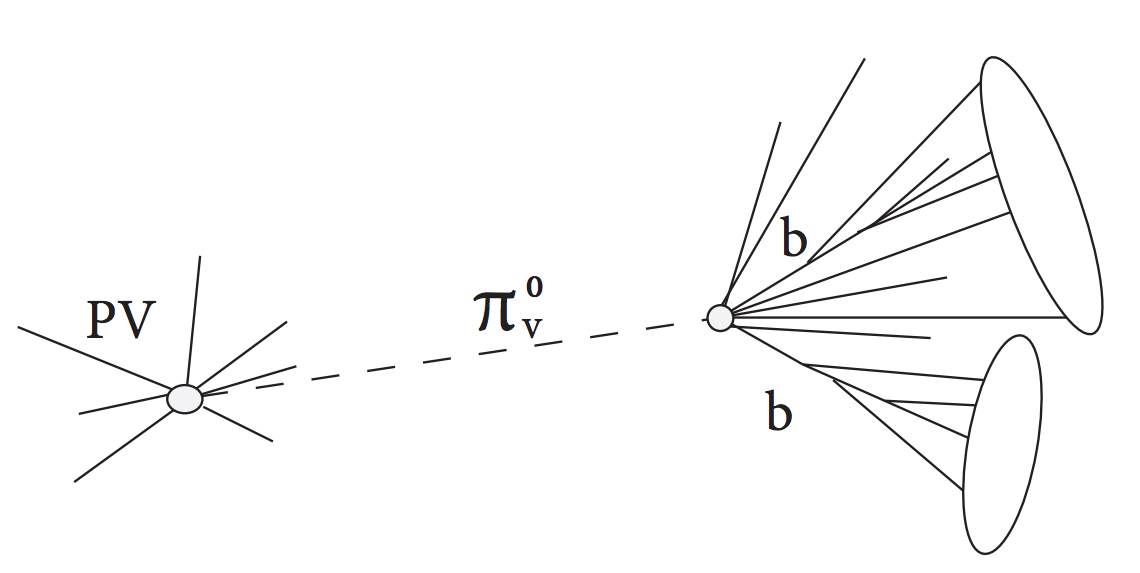}
\caption{Diagrams for the indirect (top) or direct (bottom) approach in searches. In the top diagram, a Majorana neutrino is produced off-shell in a $D^+_{(s)}$ decay to a final state with two same-sign muons (with the same diagram, the Majorana neutrino could be also produced on-shell). In the bottom one, a hidden valley pion is produced on-shell to later decay to a pair of jets.}
\label{fig:direct_indirect}
\end{figure}

These proceedings present LHCb results on the search for New Physics (NP) in Majorana neutrino and exotica related analyses. Searches at LHCb can be performed both through the indirect or direct approach. The indirect approach implies searching for new virtual particles that appear in the Feynman diagrams. This has the advantage of being potentially sensitive to higher mass scales than those accessible by direct searches, since the new particles are produced virtually, although very precise measurements are required in this case. The direct approach is somehow the more classical one, in which new particles are directly searched for in some final state. This is the strategy more typically followed by ATLAS or CMS. The indirect approach is the one followed partly for the Majorana neutrino searches, through decays of heavy mesons to final states with two same sign leptons. For the direct searches, LHCb has the advantage of its unique phase space coverage. The direct approach is the strategy followed for on-shell Majorana neutrinos and exotica searches. FIG. \ref{fig:direct_indirect} shows examples of diagrams for both direct and indirect searches.

\section[Searches for Majorana neutrinos]{Searches for Majorana neutrinos}

\subsection*{Majorana neutrinos at LHCb}
The LHC collisions produce $\sim$15 kHz $b\bar{b}$ pairs and $\sim$300 kHz $c\bar{c}$ pairs in the LHCb detector \cite{tobias}, which allows to set very stringent limits on rare $B$ and $D$ decays. In particular, this makes possible the search for off-shell and on-shell Majorana neutrinos, with final states in which two same-sign muons are present. The searches profit from the aforementioned excellent mass and lifetime resolutions of LHCb.

\begin{figure}[htbp]
\centering
\includegraphics[width=0.45\textwidth]{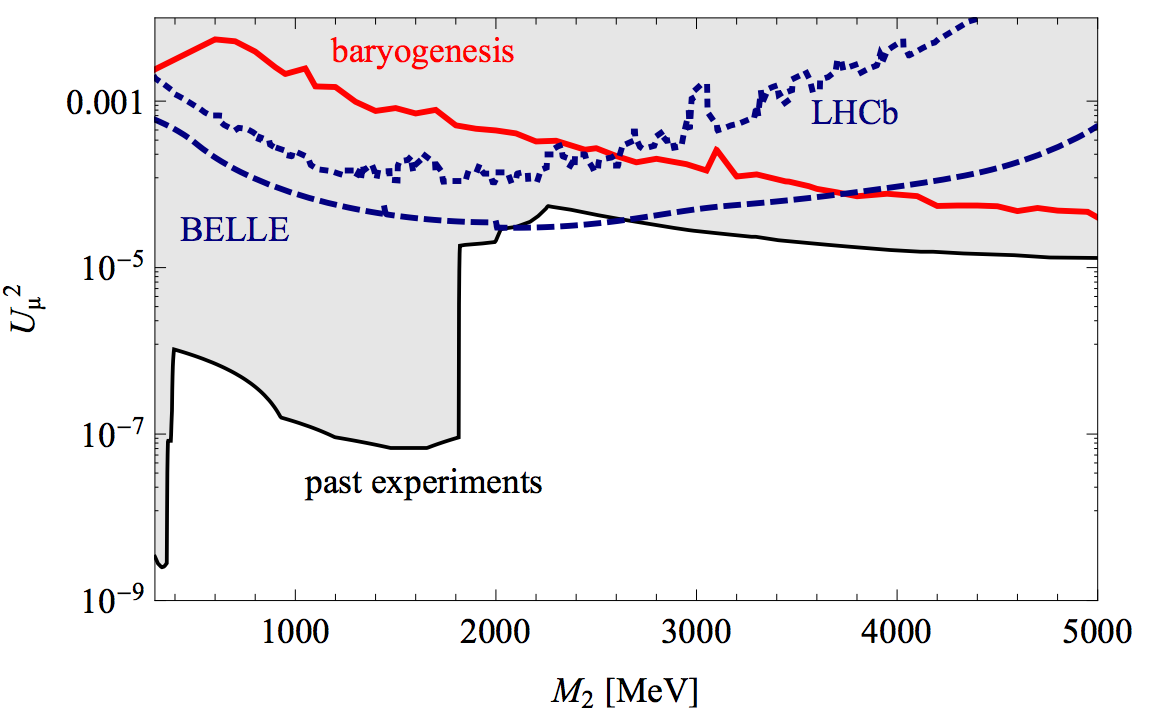}
\caption{Allowed parameter space for the model presented in reference \cite{theory}. The x-axis shows the mass of the Majorana neutrino and the y-axis the coupling of the neutrino to muons squared. Constraints by the different experiments are also shown. The Belle search can be found in reference \cite{belle_neutrinos}, although the theoretical reference used to obtain the limit on the coupling is different to that of LHCb. The red line shows the constraint from the observed baryon asymmetry of the Universe.}
\label{fig:maj_theory}
\end{figure}

Searches in $B$ and $D$ decays are complementary to other searches, such as those looking for neutrino-less double $\beta$ decay \cite{doublebeta}. In this case, LHCb searches are useful to study the coupling of Majorana neutrinos to muons. There exist specific NP models with Majorana neutrinos that can be constrained by LHCb results. An example can be found in reference \cite{theory}, where a type-I seesaw model with three right-handed neutrinos is presented. FIG. \ref{fig:maj_theory} shows the fraction of phase space constrained by LHCb results.

The main results of LHCb in the search for Majorana neutrinos are summarized below.

\subsection*{\texorpdfstring{$\mathbf{B^\pm \to h^\mp \mu^\pm \mu^\pm}$}{B+->h-mu+mu+}}

\begin{figure}[htbp]
\includegraphics[width=0.45\textwidth]{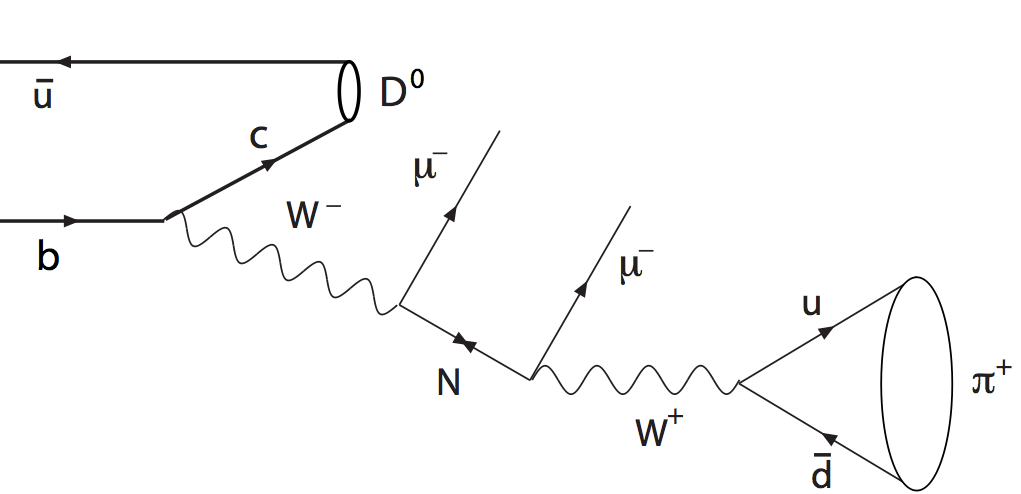}

\vspace{.5cm}
\hspace{0.3cm} \includegraphics[width=0.45\textwidth]{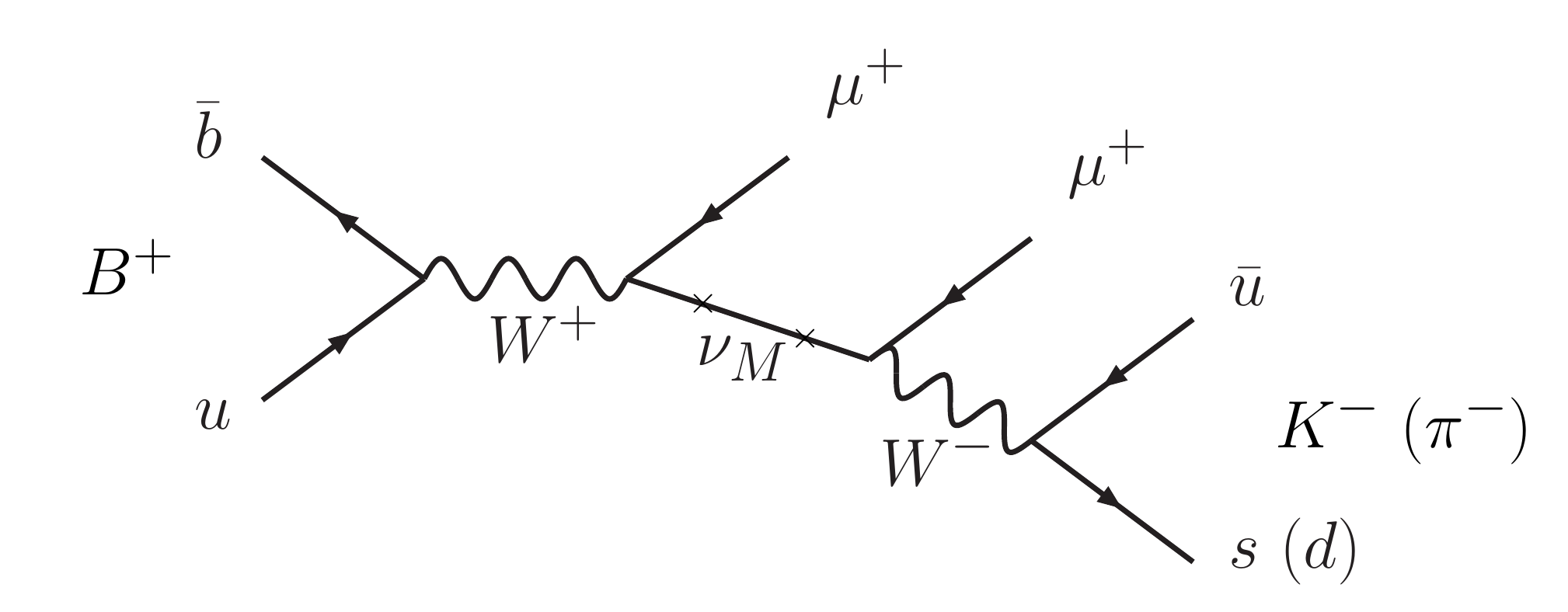}
\caption{Examples of Feynman diagrams for the search for Majorana neutrinos in $B^+$ decays.}
\label{fig:b_maj}
\end{figure}

LHCb results in $B^\pm$ decays probe a wide range of Majorana neutrino masses and lifetimes. The searches are of the type $B^\pm \to h^\mp \mu^\pm \mu^\pm$ and the results are gathered in three different papers, with different final states and LHCb datasets. FIG. \ref{fig:b_maj} shows examples of the Feynman diagrams for these decays. The final states and associated papers are:

\begin{itemize}
\item $h^\mp = K^\mp$ or $\pi^\mp$, with $\sim$36 pb$^{-1}$ ($\sqrt{s}=$7 TeV) \cite{maj1}
\item $h^\mp = D^\mp$, $D^{\ast \mp}$, $D^\mp_s$ and $D^0 \pi^\mp$, with $\sim$40 pb$^{-1}$ ($\sqrt{s}=$7 TeV) \cite{maj2}
\item $h^\mp = \pi^\mp$, with 3.0 fb$^{-1}$ ($\sqrt{s}=$7 TeV + $\sqrt{s}=$8 TeV)  \cite{maj3}
\end{itemize}

\begin{figure}[htbp]

\vspace{.25cm}
\includegraphics[width=.48\textwidth]{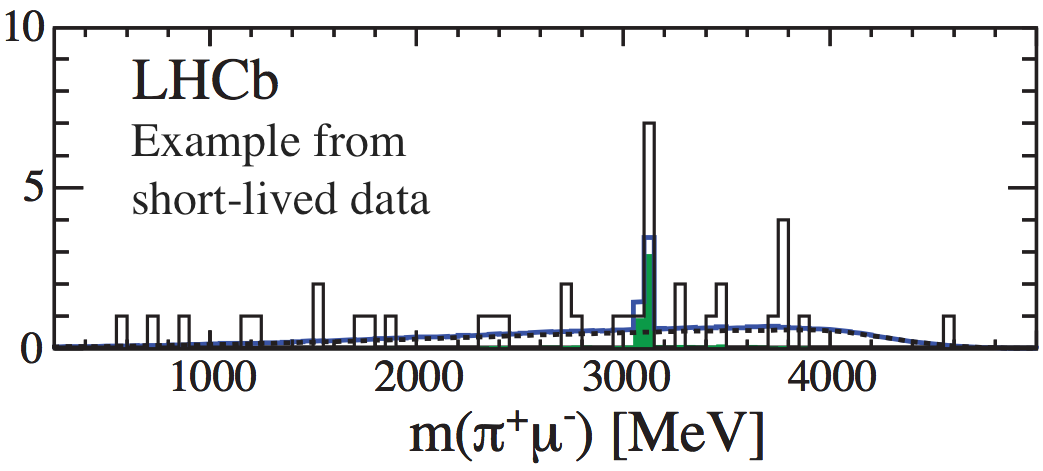}

\vspace{.25cm}

\includegraphics[width=.5\textwidth]{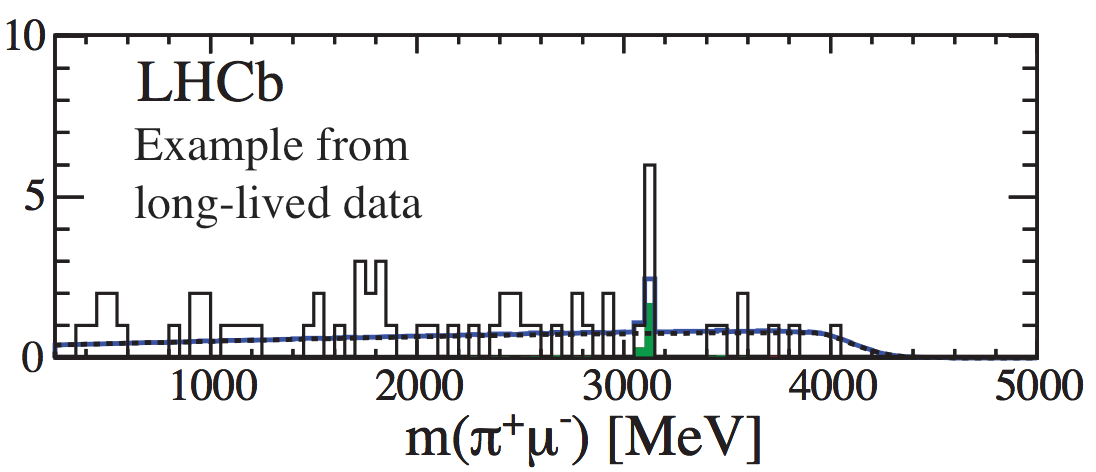}

\vspace{.25cm}
\hspace{0.5cm}
\includegraphics[width=.45\textwidth]{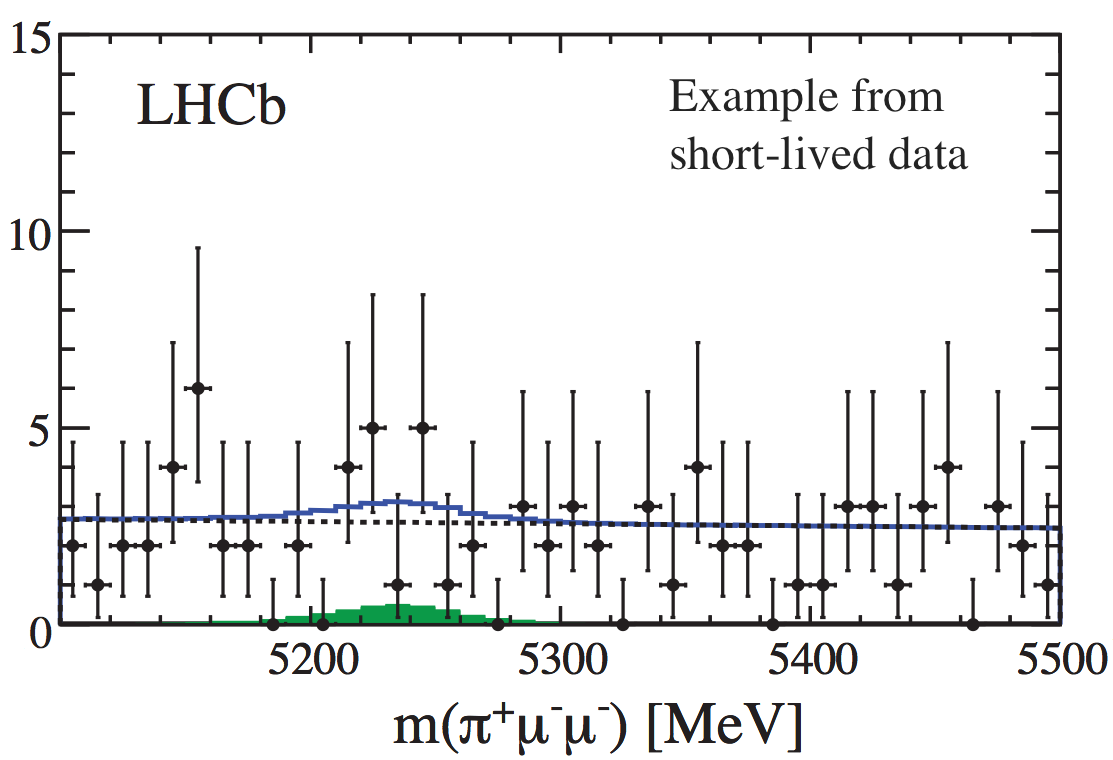}

\caption{Mass spectra in data for Majorana neutrino searches in reference \cite{maj3}. Top and medium, $\pi^+\mu^-$ for a 2$\sigma$ window around the $B^+$ mass ($\sigma$ being the mass resolution);  bottom, $\pi^+\mu^-\mu^- $. The expected contribution from charmonium backgrounds is shown in green. For the $\pi^+\mu^-$ mass spectra, top plot shows the case in which the Majorana neutrino is short-lived and medium plot the case in which it is long-lived. The $\pi^+\mu^-\mu^-$ mass spectra is only shown for the short-lived case. For the long-lived (short-lived) case, the Majorana neutrino decay vertex is (not) detached from the $B^+$ decay vertex.}
\label{fig:maj_spec}
\end{figure}

For these searches, the Majorana neutrino is generally assumed to be short-lived, i.e., not detached from the $B^+$ decay vertex. However, in reference \cite{maj3}, the possibility of a long-lived Majorana neutrino is also considered. In order to convert the yield into a branching fraction, a normalization channel, with a well-known branching fraction, is chosen. $B^+ \rightarrow J/\Psi K^+$ (with $J/\Psi \to \mu^+\mu^-$) is the normalization channel chosen for the 3-body final state channels and  $B^+ \rightarrow \Psi (2S) K^+$ (with $\Psi (2S) \to J/\Psi \pi^+ \pi^-$ and $J/\Psi \to \mu^+\mu^-$) for the 5-body final state channels. The dominant background for this analysis is that of charmonium decays, in which one of the final state particles is misidentified. The expected contribution from this background is estimated from data. As an example, FIG. \ref{fig:maj_spec} shows the relevant mass spectra for the searches performed in reference \cite{maj3}.

\begin{figure}
\includegraphics[width=0.45\textwidth]{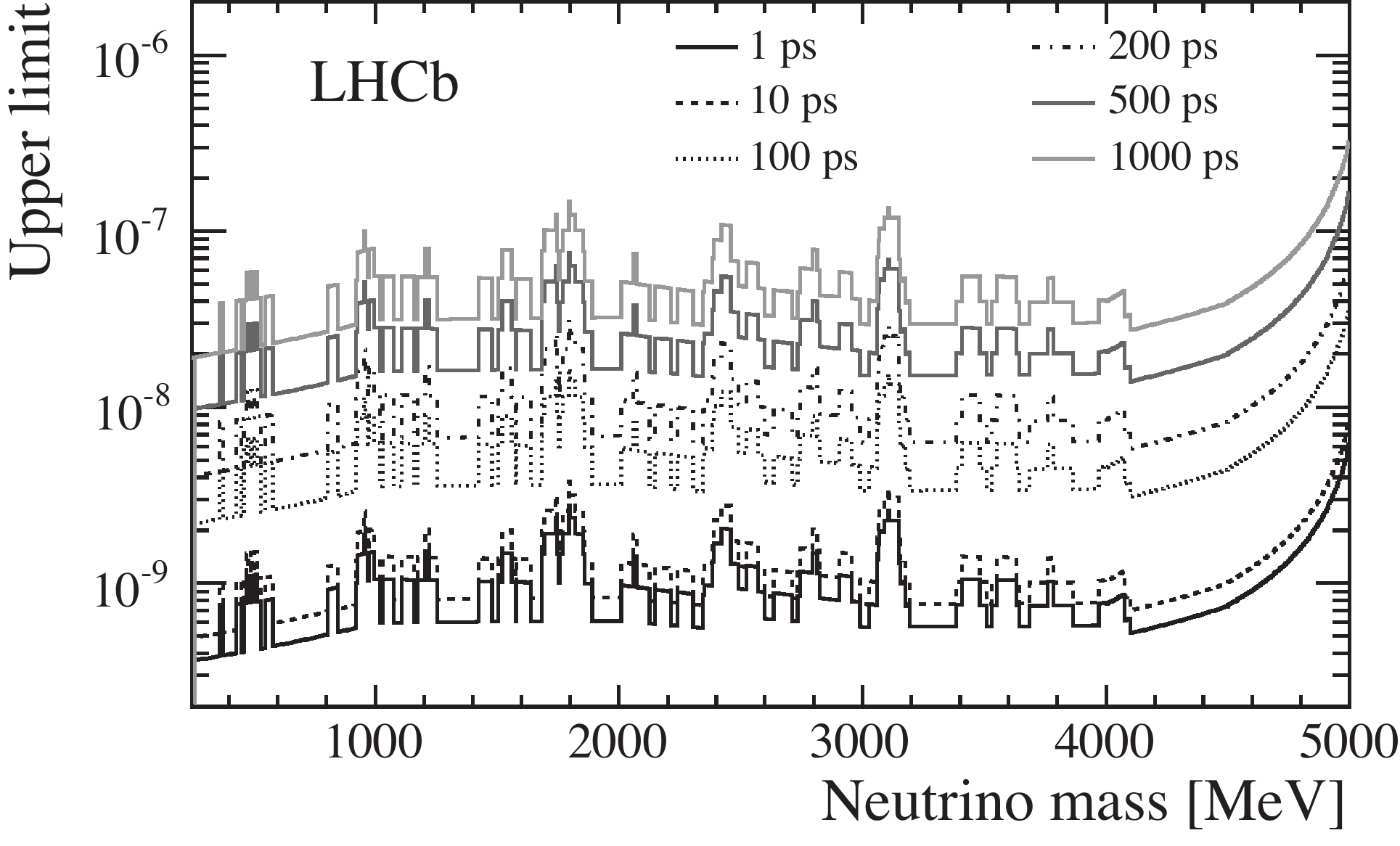}

\vspace{0.5cm}
\includegraphics[width=0.45\textwidth]{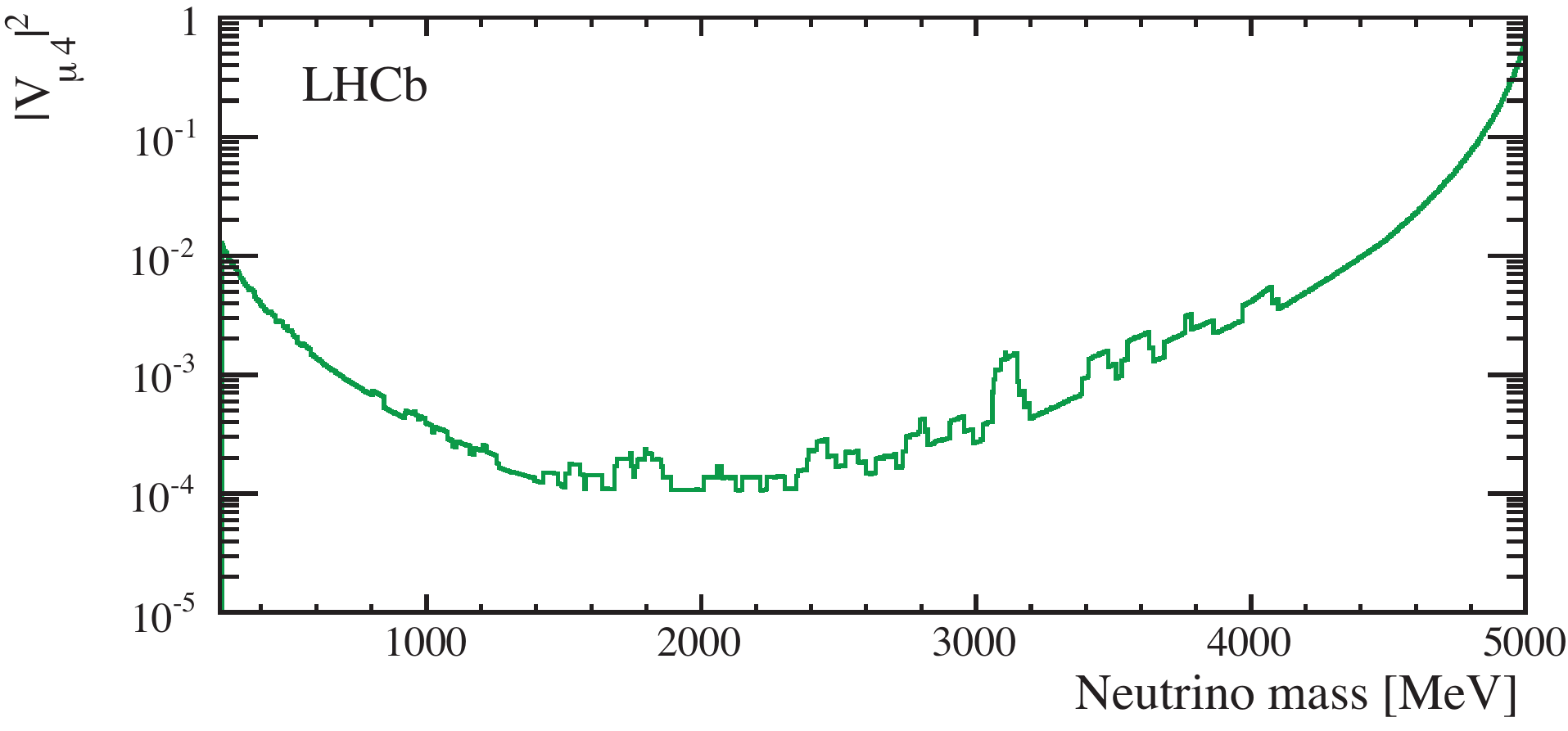}
\caption{Main results from the Majorana neutrino search in reference \cite{maj3}. Top, upper limit on the branching fraction  as a function of mass and lifetime of the neutrino. Bottom, limits on fourth generation couplings, $|V_{\mu4}|^2$, as a function of neutrino mass. The limits in both cases are at 95\% CL.}
\label{fig:maj_limits}
\end{figure}

No signal is observed in any of the searches, so upper limits on the branching fraction are set. For the searches of reference \cite{maj3}, upper limits are set by scanning across in 5 MeV/c$^2$ steps of neutrino mass from 250 to 5000 MeV/c$^2$ with varying mass resolution. For long lived neutrinos, the scan is done both in terms of the mass and lifetime of the neutrino. FIG. \ref{fig:maj_limits} shows these upper limits and also the interpretation of the limit as bounds on the couplings of a fourth generation lepton to muons. The explanation of how this coupling is calculated is given in reference \cite{maj3}. The upper limits on the branching fraction for short-lived neutrinos are calculated from the average detection efficiency using the $\rm CL_S$ method. The limits for all the decay modes under scrutiny are shown in TABLE \ref{tab:maj}. All these limits are world's best, and improve the previous limits by as much as factor of $\sim$ 100.

\begin{table}
\centering
\begin{tabular}{|c|c|}
\hline 
Channel & $\BR_{\rm UL}$ 95$\%$ CL \\ \hline
$B^\pm \to K^\mp \mu^\pm \mu^\pm$ & $5.4 \times 10^{-8}$ \\
$B^\pm \to D^\mp \mu^\pm \mu^\pm$ & $6.9 \times 10^{-7}$ \\  
$B^\pm \to D^{\ast \mp} \mu^\pm \mu^\pm$ & $2.4 \times 10^{-6}$ \\ 
$B^\pm \to \pi^\mp \mu^\pm \mu^\pm$ & $4.0 \times 10^{-9}$ \\
$B^\pm \to D^\mp_s \mu^\pm \mu^\pm$ & $5.8 \times 10^{-7}$ \\
$B^\pm \to D^0 \pi^\mp \mu^\pm \mu^\pm$ & $1.5 \times 10^{-6}$ \\ \hline
\end{tabular} 
\caption{Summary of upper limits on the branching fraction of different channels in Majorana neutrino searches in references \cite{maj1,maj2,maj3}.} 
\label{tab:maj}
\end{table}

\subsection*{\texorpdfstring{$ \mathbf{D^\pm_{(s)} \to \pi^\mp \mu^\pm \mu^\pm}$}{D+(s)->pi-mu+mu+}}

The physics case for this analysis is similar to the one presented for the $B$ decays. An example process was
previously given in the top diagram of FIG. \ref{fig:direct_indirect}. This search is performed using the the LHCb 2011 data sample, i.e., 1 fb$^{-1}$ at $\sqrt{s}=$7 TeV \cite{maj4}. The Majorana neutrino is considered to be short-lived.

\begin{figure}
\centering
\includegraphics[width=0.45\textwidth]{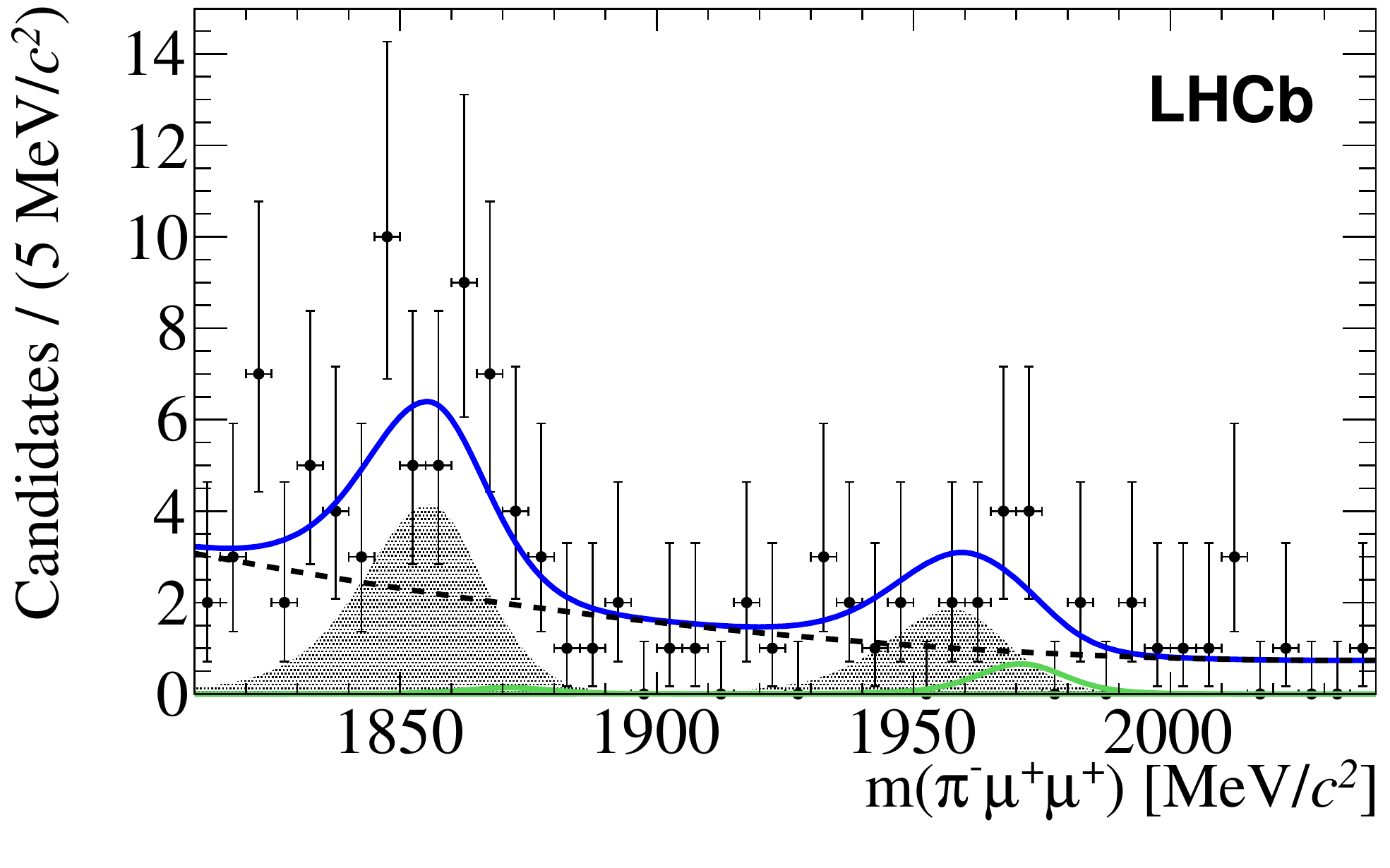}
\caption{$\pi^-\mu^+\mu^+$ mass spectra for the search in reference \cite{maj4}. The expected peaking background from $D^+_{(s)} \to \pi^+\pi^+\pi^-$ is represented by the filled grey histograms.}
\label{fig:d_maj}
\end{figure}

For this analysis, the normalization is done to the $D^+_{(s)} \to \phi(\mu^+ \mu^-)\pi^+$ decay. Furthermore, the candidates are classified using PID cuts and a BDT that included kinematic and geometrical variables. The important peaking background from $D^+_{(s)} \to \pi^+\pi^+\pi^-$ decays is also taken into account, with the expected yield also measured from data. FIG. \ref{fig:d_maj} shows the invariant mass spectra obtained in data, with the relevant peaking background expected.

No significant signal is found in this search, so upper limits on the corresponding branching fractions are set. For this, a fit in bins of $m(\pi^-\mu^+)$ is performed in order to improve the statistical significance. The results, shown in TABLE \ref{tab:dmaj_res}, correspond to an improvement by a factor of 50 with respect to previous limits.

\begin{table}
\centering
\begin{tabular}{|c|c|}
\hline 
Channel & $\BR_{\rm UL}$ 95$\%$ CL \\ \hline
$D^\pm \to \pi^\mp \mu^\pm \mu^\pm$ & $2.5 \times 10^{-8}$ \\
$D^\pm_s \to \pi^\mp \mu^\pm \mu^\pm$ & $1.4 \times 10^{-7}$ \\ \hline
\end{tabular} 
\caption{Upper limit on the branching fractions of the decays searched for in reference \cite{maj4}.}
\label{tab:dmaj_res}
\end{table}

\section[Direct searches for exotics]{Direct searches for exotics}

\subsection*{\texorpdfstring{Limits on $H^0 \to \tau^+\tau^-$ production}{Limits on H0->tautau production}}

\begin{figure}[htbp]
\centering
\includegraphics[width=0.4\textwidth]{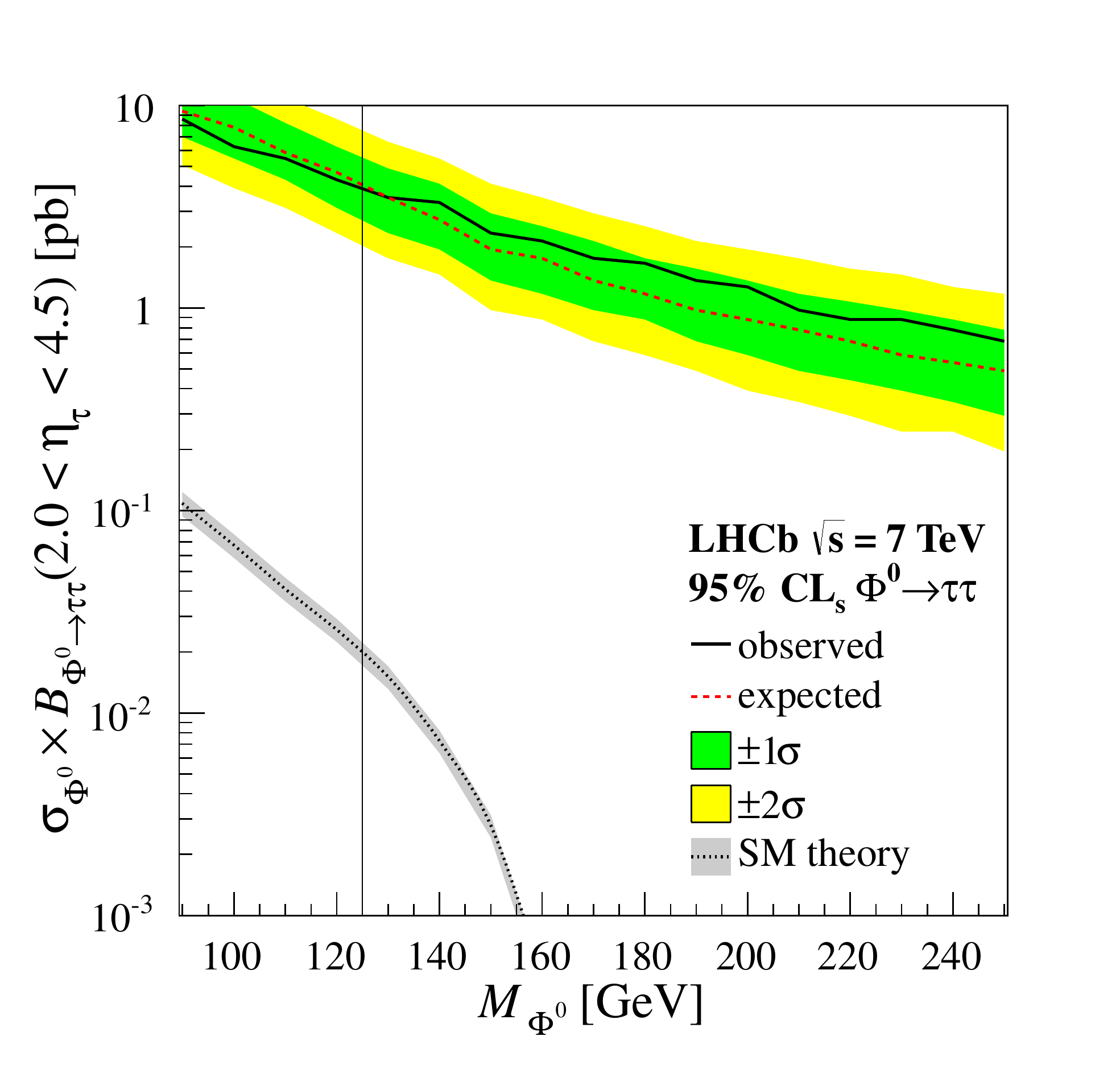}

\hspace{0.15cm}\includegraphics[width=0.4\textwidth]{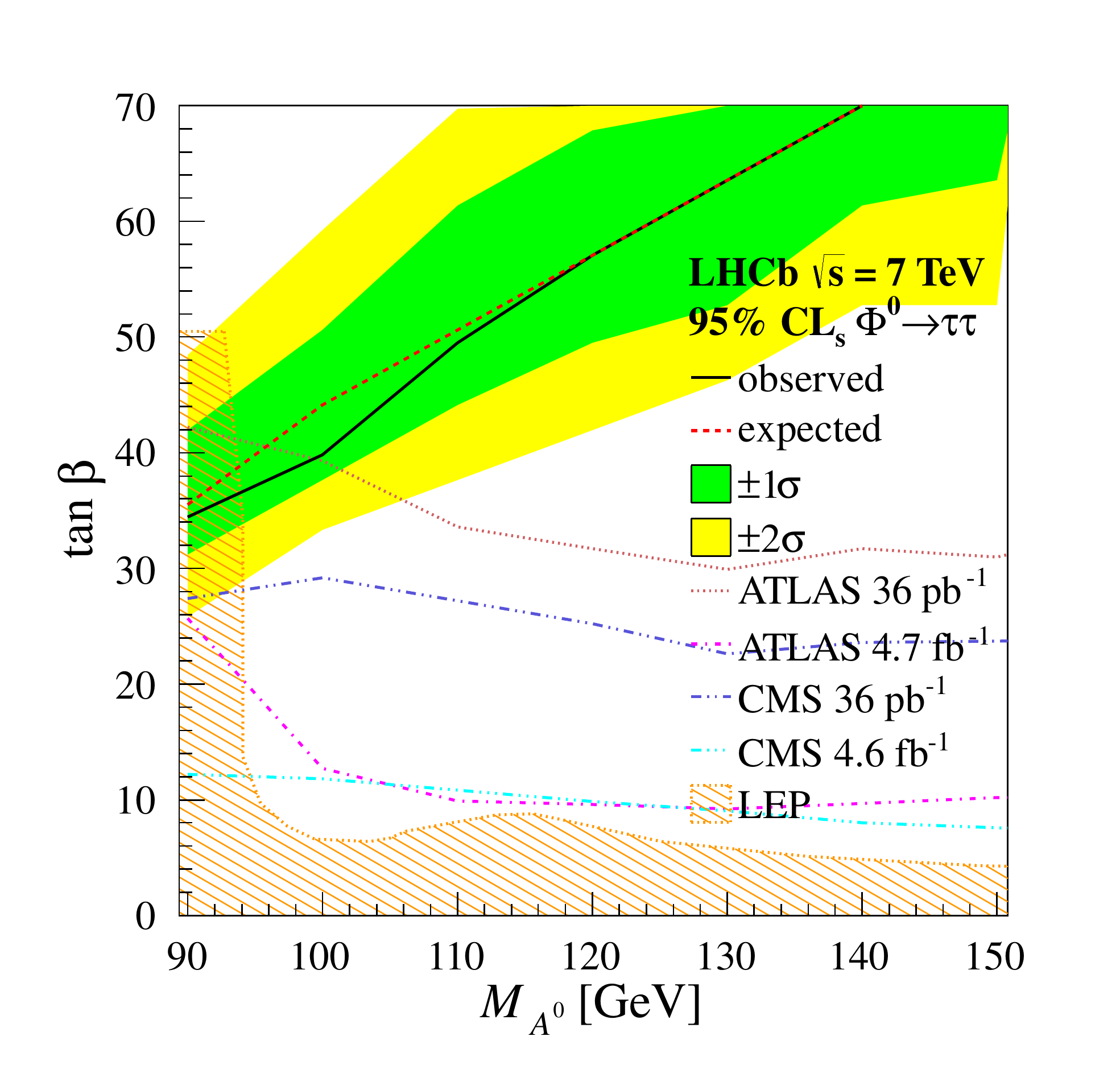}
\caption{Results on the $H^0\rightarrow \tau^+\tau^-$ search at LHCb \cite{higgstautau}. All limits are at 95\% CL, set using the $\rm CL_S$ method. Top, model independent limit in terms of 
$\sigma_H \times BR(H^0\rightarrow \tau^+\tau^-)$ (for the SM prediction, see \cite{htautau_SM1,htautau_SM2}). Bottom, limit compared to ATLAS \cite{htautau_atlas1,htautau_atlas2}, CMS \cite{htautau_cms1,htautau_cms2} and LEP \cite{htautau_lep} in the $m(h^0)_{max}$ scenario \cite{maxh0}. }
\label{fig:htautau}
\end{figure}

This is the first LHCb paper concerning a search for the Higgs boson \cite{higgstautau}. 
No excess is found in the 2011 dataset, so upper limits on the production are set using the CL$_{\rm S}$ method. This is done both in a model independent way (as a function of the Higgs boson mass, $m_H$) and in one particular realization of the Minimal Supersymmetrical Standard Model (MSSM).
FIG. \ref{fig:htautau} shows the main results of this search.

\subsection*{Higgs decays to long-lived particles}

LHCb carries out a search for long-lived massive particles decaying from a Standard Model-like Higgs boson. The search is performed using using 0.62 fb$^{-1}$ of the 2011 LHCb dataset \cite{detached}. 
In this case, the full 2011 dataset is not used since the most efficient triggers were only operated from summer 2011 on. Several NP models predict this kind of particles. Examples are:
\begin{itemize}
\item  SUSY models with Baryon number Violation (BV), 
  $h^0 \to \tilde{\chi}^0_1 \tilde{\chi}^0_1$, with $\tilde{\chi}^0_1$ being a long-lived neutralino and $\tilde{\chi}^0_1 \rightarrow$ 3 quarks \cite{rparity}.
\item Some Hidden Valley (HV) models, 
$h^0 \to \pi^0_V \pi^0_V$, and $\pi^0_V$ decays to a pair of displaced $b-$quarks \cite{hidden}.
\end{itemize}

\begin{figure}[htbp]
\includegraphics[width=0.45\textwidth]{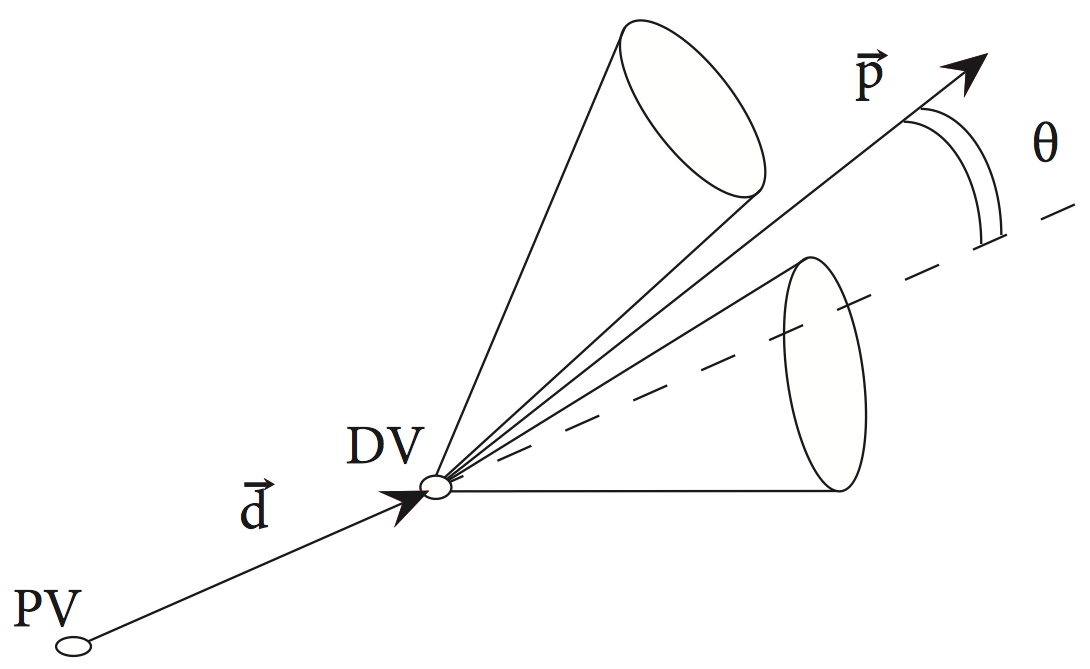}

\vspace{1cm}
\hspace{0.75cm}\includegraphics[width=0.45\textwidth]{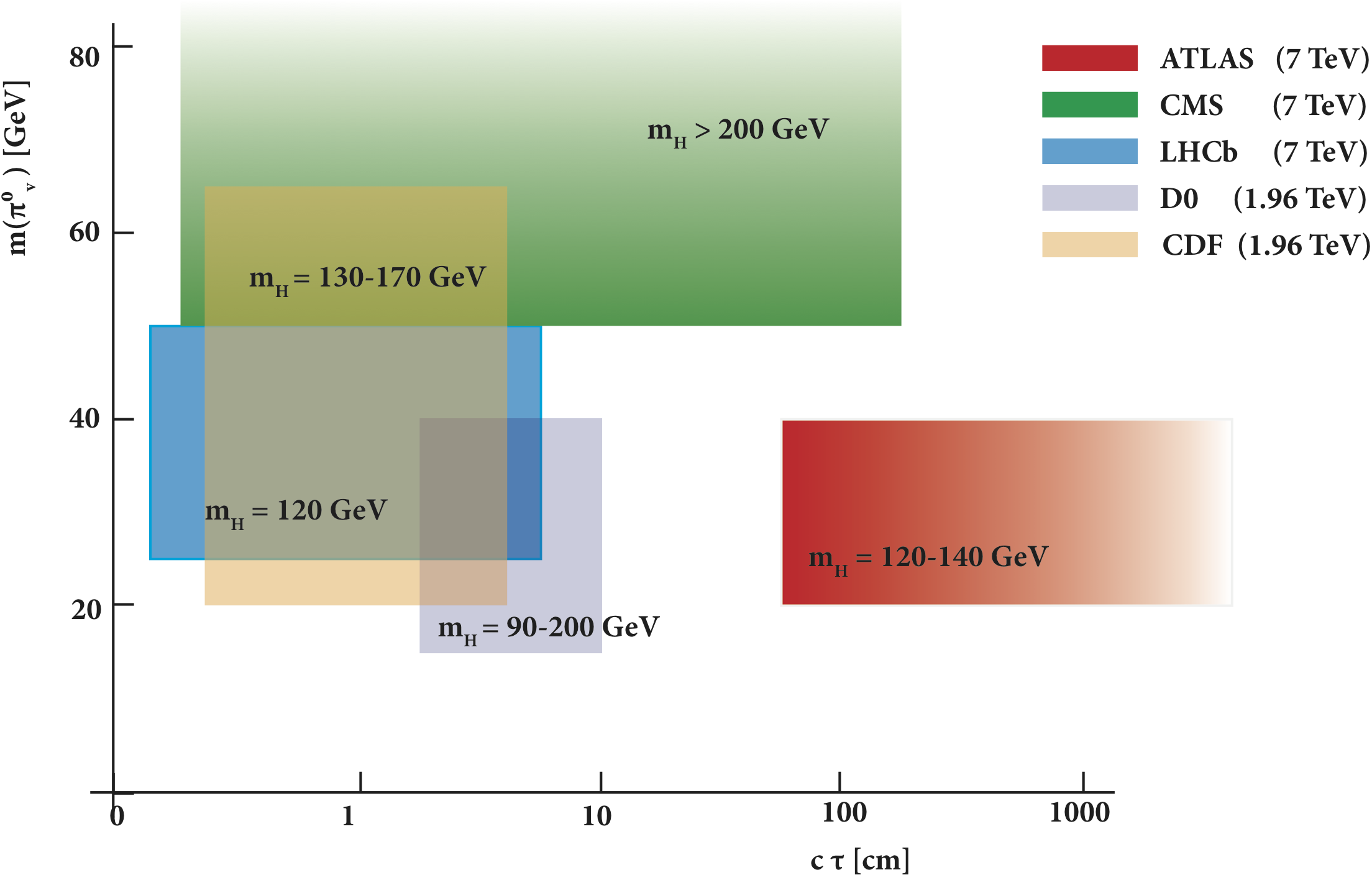}
\caption{LHCb on $H^0$ decays to long-lived particles. Top, experimental signature. Bottom, LHCb parameter space coverage in HV models compared to other experiments: x-axis corresponds to the long-lived particle lifetime and y-axis to its mass.}
\label{fig:lhcb_detached}
\end{figure}

The experimental signature for this search is a displaced vertex with two associated jets. This signature specifically profits the LHCb advantage on aspects such as vertexing. Furthermore, it should be noted that, concerning the parameter space coverage, the contribution from LHCb is particularly interesting and complementary to that of other experiments. FIG. \ref{fig:lhcb_detached} shows a diagram of the experimental signature and a plot of the LHCb complementarity in the specific case of HV models.

\begin{figure}[htbp]

\includegraphics[width=0.45\textwidth]{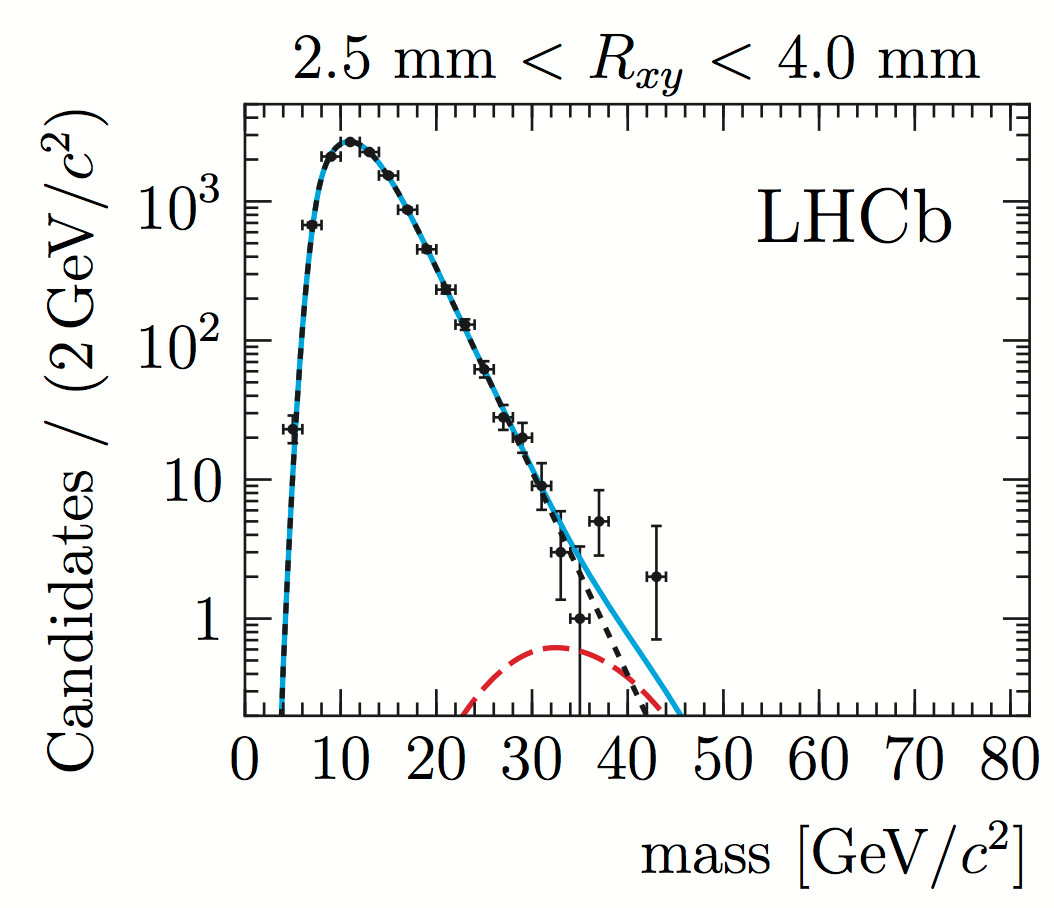}

\vspace{.5cm}
\includegraphics[width=.45\textwidth]{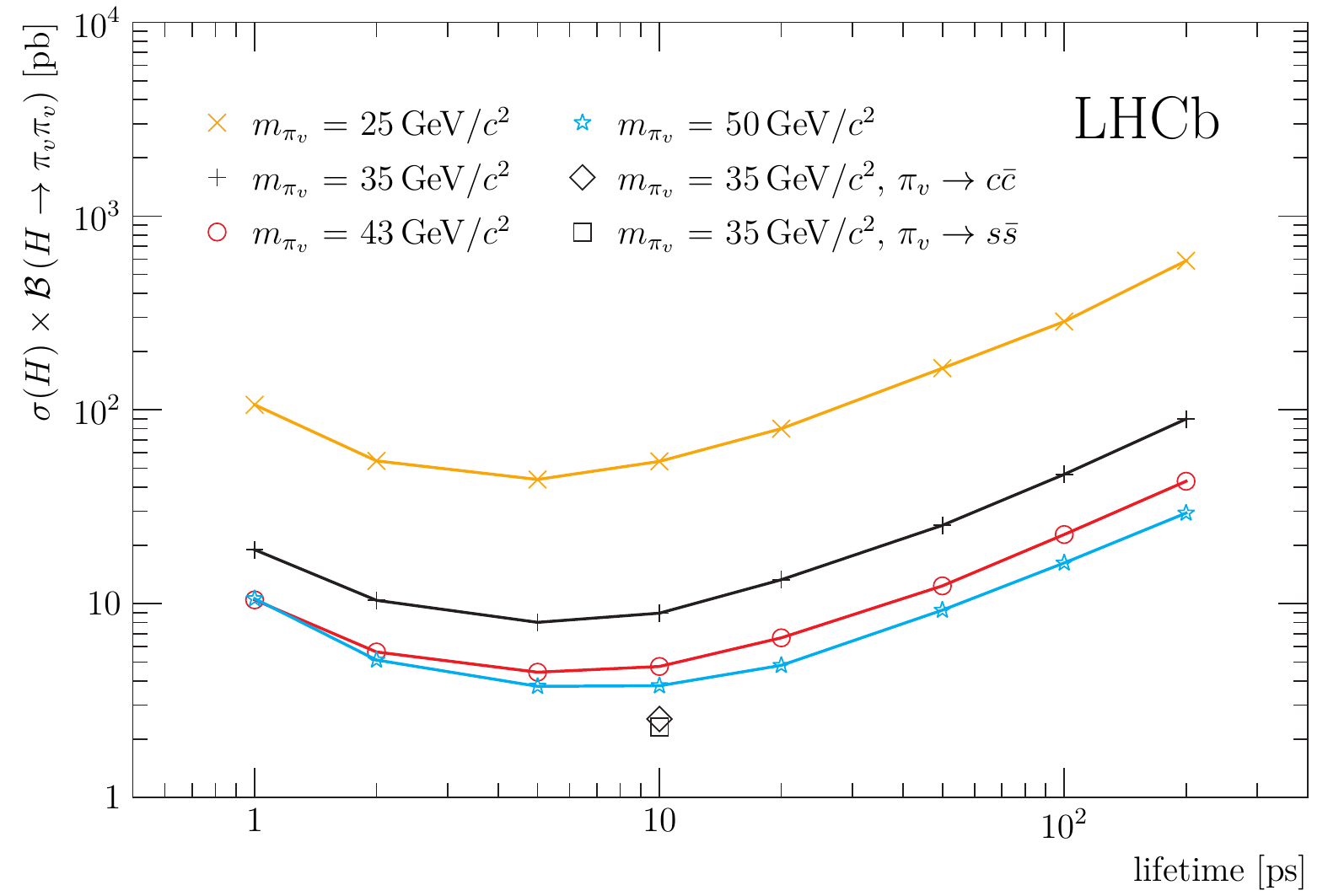}

\caption{Main results from reference \cite{detached}. Top, example from one of the fits to different radial distances. Bottom, 95\% CL upper limits on the production depending on the lifetime and and mass of the long-lived particle.}
\label{fig:detached_results}
\end{figure}

This search finds no excess above background, which was mainly composed of $b \bar{b}$ pairs. Because of this, upper limits on the production are set in different regions of the NP models parameter space. For this, fits are performed for different radial distances of the long-lived particles. The radial distance works here as a proxy for the long-lived particles lifetime. FIG. \ref{fig:detached_results} shows one of these fits and the limits as a function of the long-lived particles mass and lifetime. For reference, the complementary CMS and ATLAS results can be found in references \cite{CMS-PAS-EXO-12-038,atlas_llp}, although they cover a different phase-space, so they are not directly comparable.

\subsection*{Search for long-lived heavy charged particles}

Using an innovative approach, a search for charged massive stable particles (CMSP) is performed using the LHCb detector and the 2011 and 2012 datasets \cite{stau}. 

For the analysis, the $\tilde{\tau}$ from the mGMSB model \cite{stau_th1,stau_th2,stau_th3} is used as a benchmark.

When performed at other experiments, these kinds of analyses are typically based on $dE/dx$ and time-of-flight measurements. However, LHCb developed a novel technique that is based on the Ring Imaging Cherenkov  (RICH) detector \cite{rich}. Using the information from the RICH, a Delta Log Likelihood is built to distinguish CMSPs (with lower $\beta$) from other standard stable particles. In particular, since CMSPs are slow, the
absence of corresponding light rings in the RICH is looked for. The main background for this search is Drell-Yan pair-production of $\mu^+\mu^-$.

\begin{figure}
\centering
\includegraphics[width=.45\textwidth]{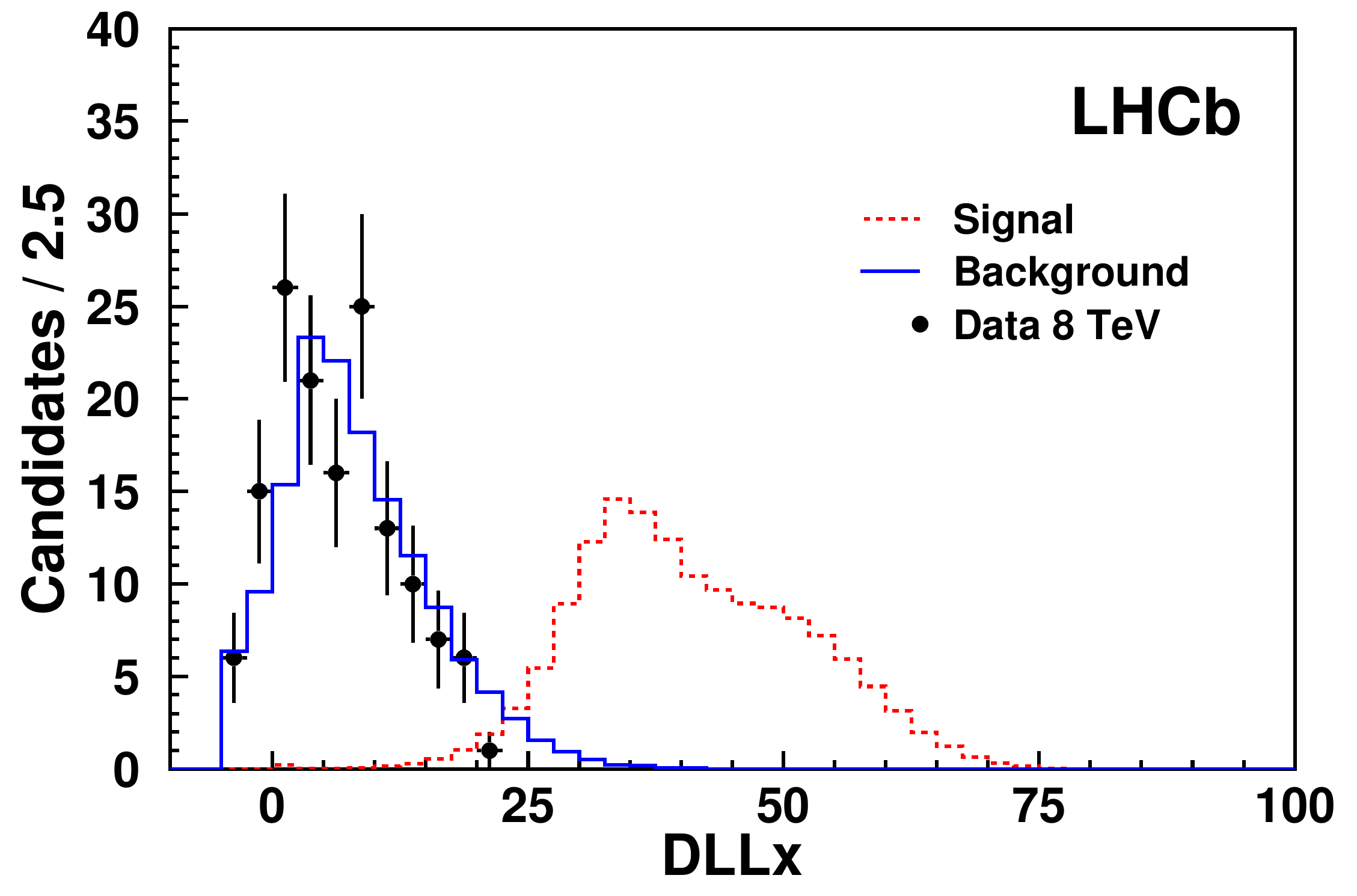} 

\vspace{.5cm}

\hspace{-.85cm}
\includegraphics[width=.45\textwidth]{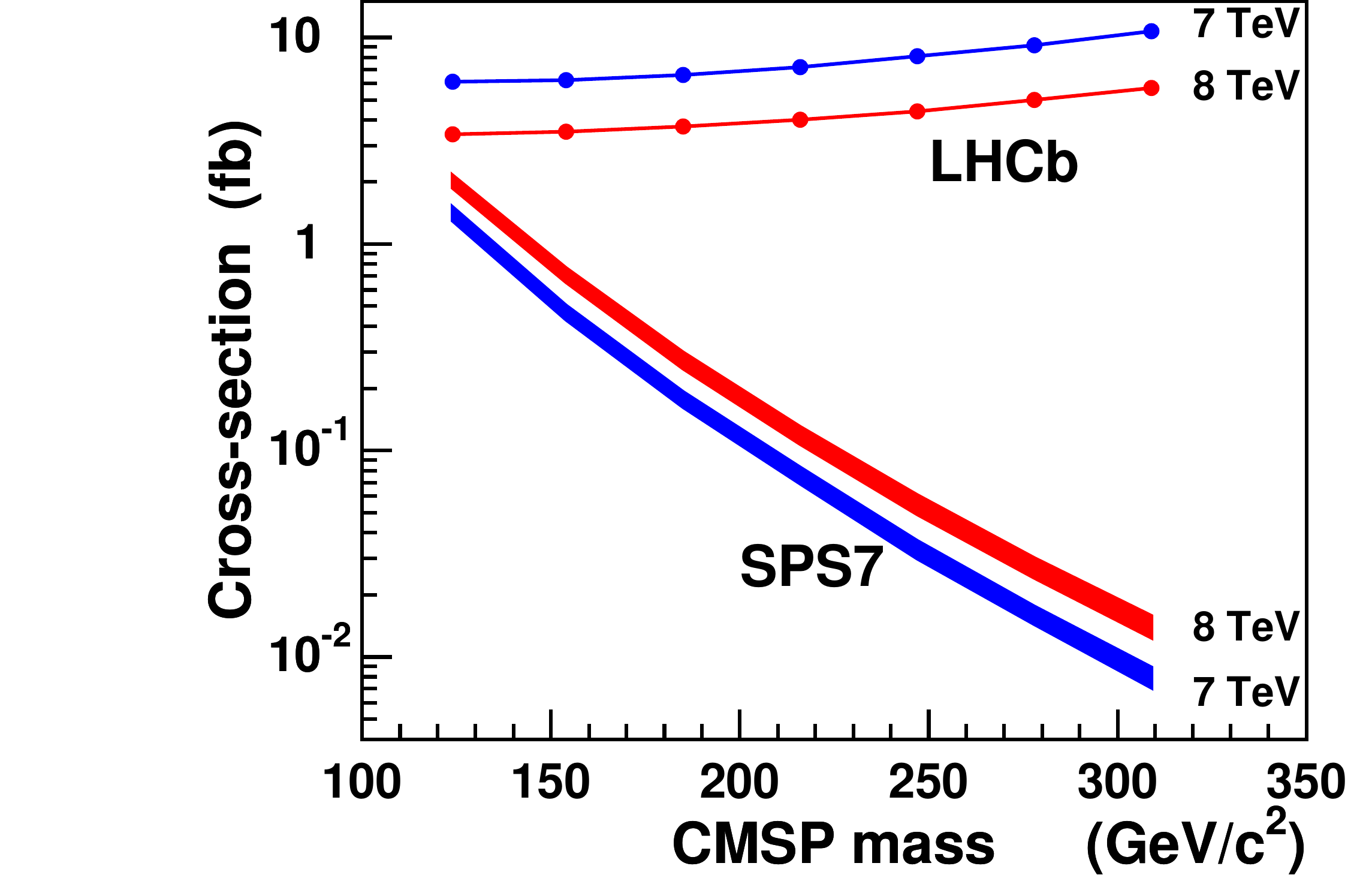} 

\caption{Results from reference \cite{stau}. Top, RICH likelihood discrimination between Drell-Yan $\mu^+\mu^-$ and a 124 GeV/c$^{2}$ long-lived particle. Bottom, 95\% CL upper limits on the production of the charged massive stable particle obtained with the Feldman-Cousins method. Theoretical cross sections in SPS7 \cite{sps7} (one of the scenarios of mGMSB) are also shown.}
\label{fig:cmsp}
\end{figure}

No signal is found in this search, so upper limits are set on its production using the Feldman-Cousins technique. FIG. \ref{fig:cmsp} shows an example of the Delta Log Likelihood built as a discriminant between CMSPs and other particles and also upper limits on the production of these particles as a function of their mass. Searches for CMSPs at ATLAS and CMS can be found in references \cite{stau_atlas,stau_cms}.

\subsection*{Searches for low mass dark bosons}

\begin{figure}[htbp]
\centering
\includegraphics[width=0.45\textwidth]{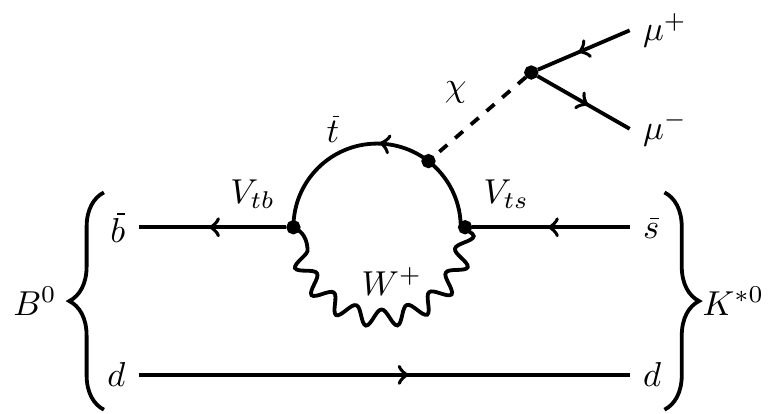}
\caption{Feynman diagram for the resonant production of a low mass dark boson in a $B^0 \to K^{\ast}\mu^+\mu^-$ decay.}
\label{fig:dark_diagram}
\end{figure}

Several theoretical models predict the existence of new particles that couple to SM particles by mixing with the Higgs. In particular inflaton, axion-like and dark matter mediator models also predict these particles to be light. LHCb performs a search for such particles \cite{b_hidden}, referred to as $\chi$, as di-muon decaying resonances in a $B$ meson decay: $B^0 \to K^{\ast}\chi(\mu^+\mu^-)$. The diagram for this decay can be found in FIG. \ref{fig:dark_diagram}.

The features of $\chi$ depend on the theoretical model considered, and in particular a wide variety of lifetimes of $\chi$ could be possible. Two benchmark models are considered in this search: inflaton models \cite{inflaton} and axion portal models \cite{axion}.

This search at LHCb relies on muon triggers, and includes the application of a multivariate selection not dependent on the $\chi$ mass or lifetime. Furthermore, in order to maximize the phase space coverage, the lifetime was factorized into prompt and detached.

\begin{figure}[htbp]
\centering
\includegraphics[width=0.48\textwidth]{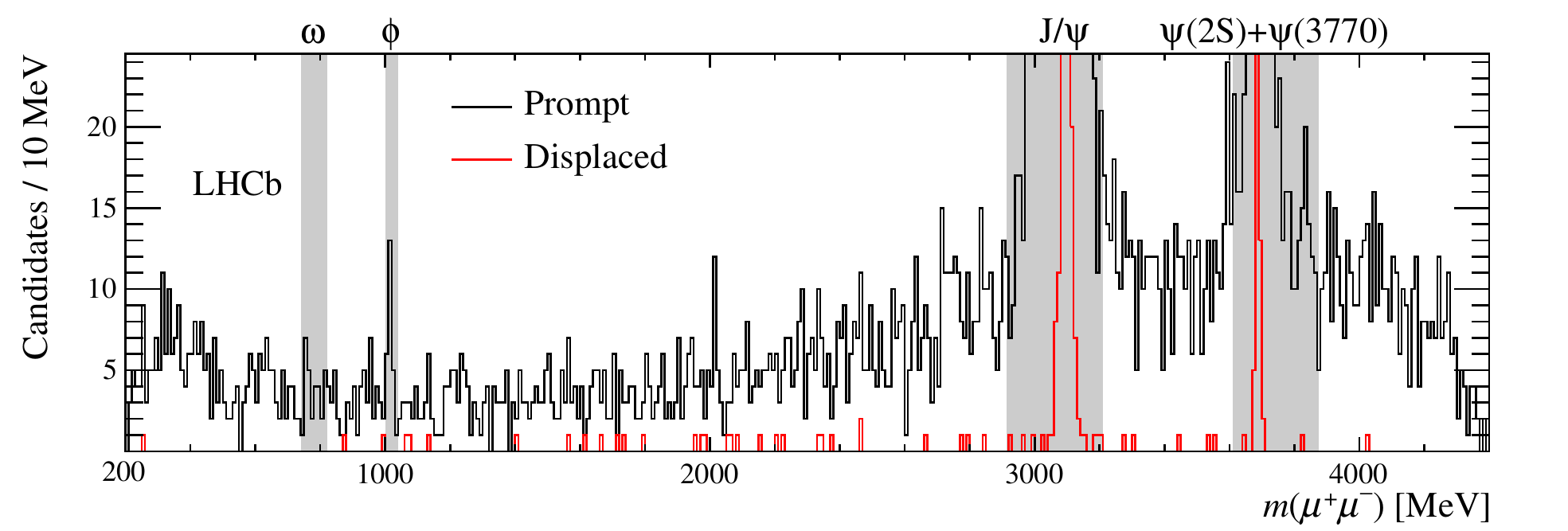}

\vspace{.5cm}
\includegraphics[width=0.48\textwidth]{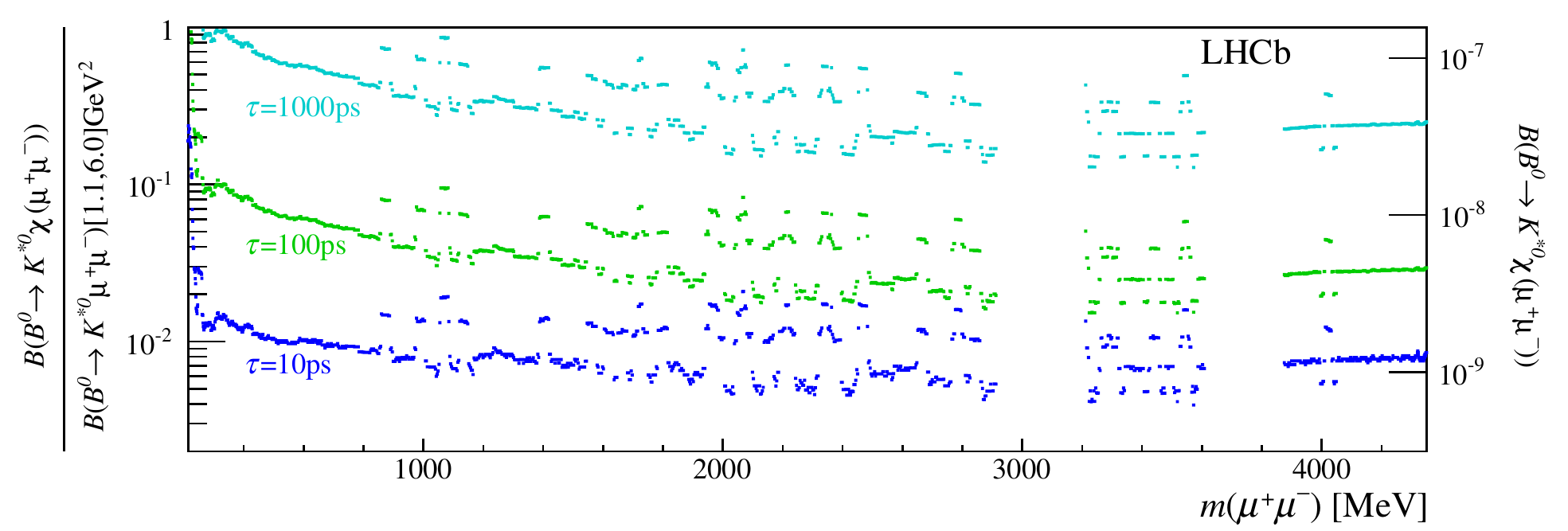}
\caption{Results of references \cite{b_hidden}. Top, $\mu^+\mu^-$ mass spectra whenever the di-muon pair is detached or not. Bottom, 95\% CL upper limits on the production as a function of the lifetime and mass of the $\chi$ resonance.}
\label{fig:dark_results}
\end{figure}

\begin{figure}[htbp]
\centering
\includegraphics[width=0.45\textwidth]{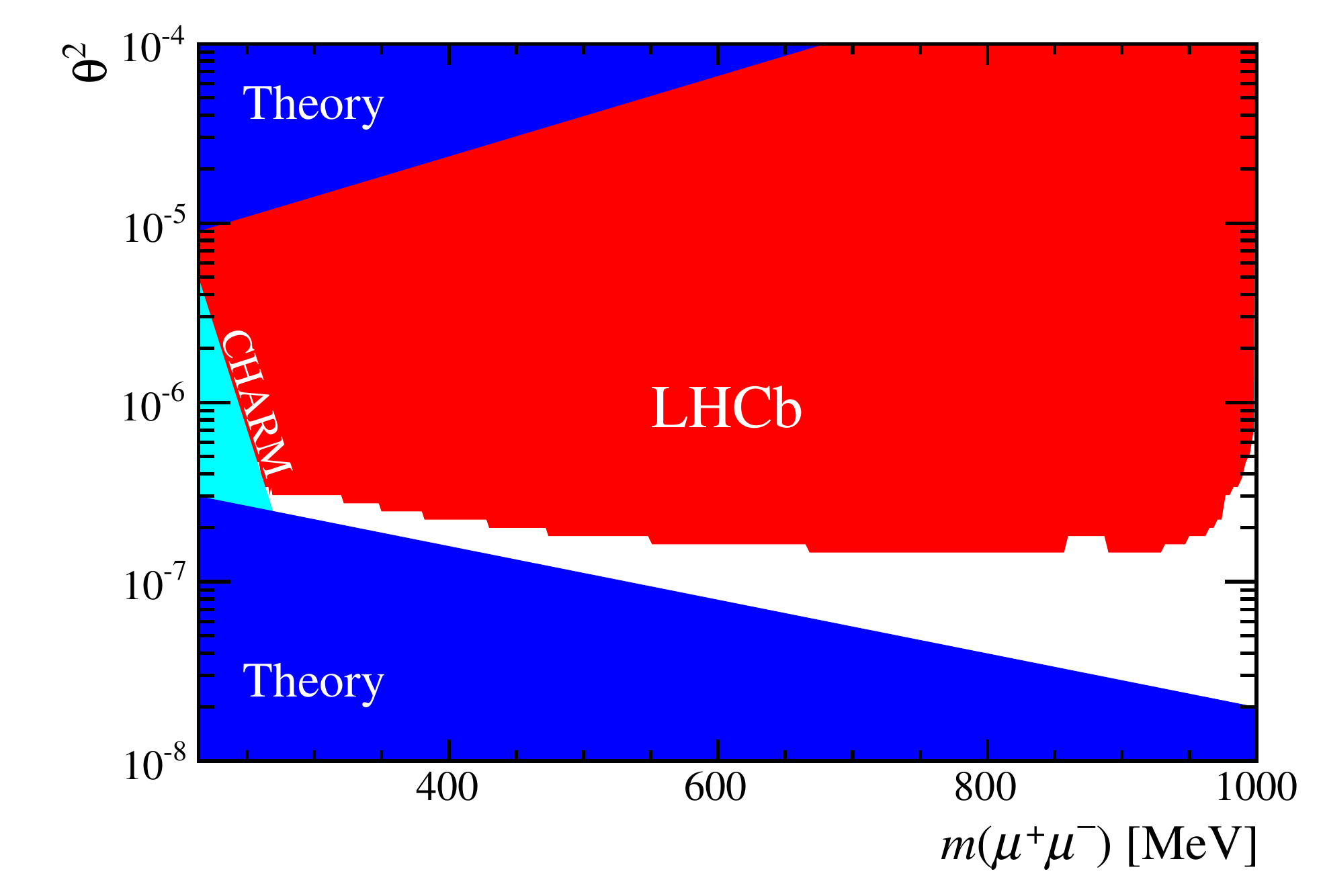}

\vspace{.5cm}
\hspace{0.3cm} \includegraphics[width=0.45\textwidth]{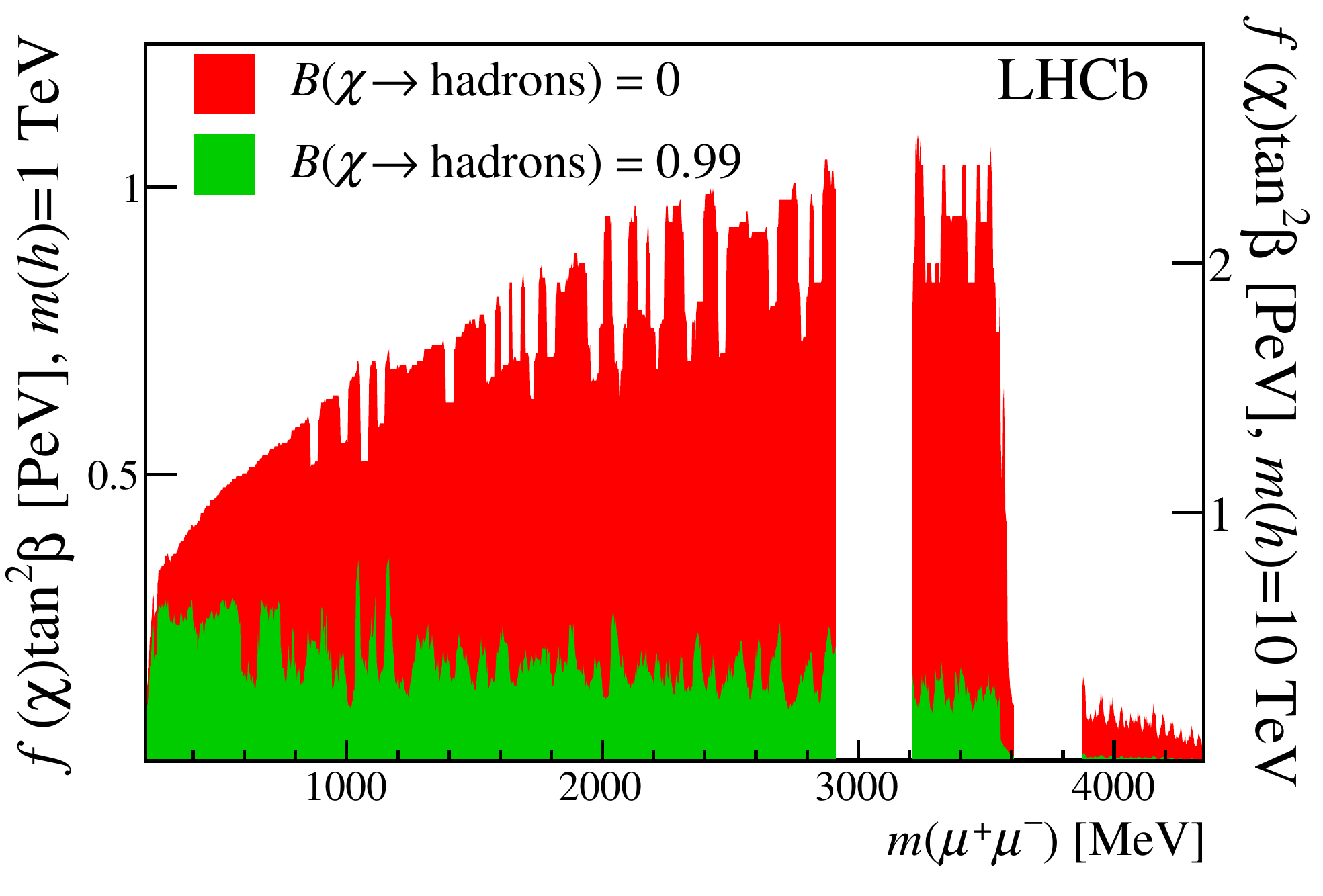}
\caption{Exclusion regions at 95\% CL in two dark matter mediator theoretical models. Top, inflaton model \cite{inflaton}. Bottom, axion model \cite{axion}. For the former, the regions excluded by the theory and by the CHARM experiment \cite{charm} are also shown.}
\label{fig:dark_inter}
\end{figure}

The search for the $\chi$ resonance is carried out by doing a scan in steps of the di-muon mass (with the $B^0$ mass constrained) and computing a test statistic, according to reference \cite{mike_dilepton}. Following this procedure, the search does not find any excess and upper limits on the the production of $\chi$ are set as a function of its lifetime and mass. FIG. \ref{fig:dark_results} shows the di-muon invariant mass spectra in this search and the resulting upper limits. Finally, FIG. \ref{fig:dark_inter} shows the parameter space constrained by this result in two specific theoretical models that predict the production of inflatons and axions \cite{inflaton,axion}.

\subsection*{\texorpdfstring{Towards $\mathbf{H^0 \to b \bar{b}}$}{Towards Higgs -> bbar}}

$H^0 \to b \bar{b}$ is interesting to study the coupling of the Higgs boson to quarks. The probability to have both $b$ quarks decaying from $H^0$ in the LHCb
acceptance is estimated to be $\sim$5\% at $\sqrt{s}=$7/8 TeV, with an expected improvement at 13 TeV. 

As a proof of the possibility of this search, LHCb has already successfully tested its jet reconstruction capacities, and also has achieved substantial progress on in $b$-jet tagging \cite{btag}. Several related analysis have been performed and are published. Some important examples are the top analysis \cite{top}, the measurements of $W$ and $Z$ + jets or $b/c$ jets \cite{wplusbc,zjets,zplusb}, and the measurement of the charge $b\bar{b}$ asymmetry \cite{bbasym}.

\section{Conclusions}

In these proceedings, LHCb has been shown to be competitive for NP searches following both direct and indirect approaches. LHCb can offer a unique coverage and complement other experiments.

Several LHCb Majorana neutrino searches have been presented. Even if these searches found no evidence, important constraints are set using both $B$ and $D$ decays to same sign muons.

Finally, several searches for exotica have been carried out at LHCb using the 2011 and 2012 datasets. Examples are the first LHCb paper on the Higgs boson: $H^0 \to \tau^+\tau^-$, the searches for long-lived particles decaying to jets and heavy long-lived stable particles as well as the study of low mass dark bosons in $B$ decays.

\providecommand{\href}[2]{#2}\begingroup\raggedright\endgroup

\end{document}